\documentclass[pdftex]{article}
\usepackage{icrctc07}

\title{Nuclear Cosmic Rays propagation in the Atmosphere}
\shorttitle{Cosmic rays propagation in atmosphere}

\authors{A. Putze $^{1}$, L. Derome $^{1}$, D. Maurin $^{2}$, M. Bu\'enerd $^{1}$}
\shortauthors{A. Putze et al.}
\afiliations{
$^1$ LPSC, Universit\'e Joseph Fourier Grenoble 1, CNRS/IN2P3 \& INPG \\ 
$^2$ LPNHE, IN2P3 - CNRS - UniversitŽs Paris VI et Paris VII}
\email{laurent.derome@lpsc.in2p3.fr}

\abstract{
The transport of the nuclear cosmic ray flux in the atmosphere is
studied and the atmospheric corrections to be applied to the
measurements are calculated. The contribution of the calculated
corrections to the accuracy of the experimental results are discussed
and evaluated over the kinetic energy range 10-10$^{3}$~GeV/n. The Boron
(B) and Carbon (C) elements system is used as a test case. It is shown
that the required corrections become largely dominant at the highest
energies investigated. The results are discussed.  }
\begin{document}
\maketitle

§§§§§§§§§§§§§§§§§§§§§§§§§§§§§§

\section{Introduction}

The direct measurement of nuclear cosmic rays (CR) by balloon borne experiments
like CREAM~\cite{2004AdSpR..33.1777S} is a precious tool to understand the
cosmic rays propagation process in the galaxy. Among the different observables,
the secondary-to-primary ratios of nuclear CR like B/C play a predominant role
because they measure directly the thickness of matter crossed by CR and
therefore tell us about their confinement in the galaxy.
%In the diffusion model, and at energy above few Gev/n, the energy
%dependence of this ratio is inversely proportional to the power law
%energy dependence of the diffusion coefficient which spectral index is
%known as the diffusion index. The high energy measurement of the
%primary-over-secondary ratio allows to directly access the spectral
%index.
In addition, the CR flux entering the atmosphere undergoes nuclear
fragmentation by interaction with the atmospheric nuclei, producing
more secondaries and therefore enhancing the measured secondary-to-primary ratios.

At high energy, the confinement of cosmic rays in the galaxy becomes less and
less efficient. The thickness of matter crossed in the galaxy is therefore
decreasing with energy whereas the thickness of matter crossed in the atmosphere
overhead the detector remains constant with energy. The production of secondaries
in the atmosphere is then becoming dominant at high energy.

%  and so for the flux of secondary
%nuclei produces in the atmosphere.

%for a given nucleus generates a significant contribution to the
%measured galactic flux at TOA for the same isotopes. For the secondary
%over primary ratios like B/C,
%The CR flux entering the atmosphere undergoes nuclear fragmentation by
%interaction with the atmosphere nuclei. These reactions, undergone by
%the galactic flux to be measured, in the crossing of the atmosphere
%layer overhead the balloon, deplete the incident flux for a given
%nucleus and correspondingly populates the flux of lighter nuclei. In
%this complex process, the produced flux of secondary nuclei for a
%given nucleus generates a significant contribution to the measured
%galactic flux at TOA for the same isotopes. For the secondary over
%primary ratios like B/C, the softer the distribution, the larger the
%effect is expected to be in the upper range of the energy spectrum.

The purpose of this work is to propose a framework to compute the evolution of
the nuclear cosmic ray composition in the atmosphere, taking into account the
absorption and fragmentation processes. It can be used to reconstruct the top of
atmosphere (TOA) flux from the value measured by a detector in the
atmosphere. The effects of statistical and systematic errors on this
reconstruction are also carefully studied.

%calculation is to evaluate the flux of some
%relevant nuclear species at a given altitude, i.e., for a given
%grammage of crossed atmospheric matter, by solving a standard
%propagation equation.
%
\subsection{Transport model}\label{TRANS}
The weighted slab model (WSM) provides a particularly appropriate
framework for the calculations since in this approach, the variable of
the transport equation is the amount of matter crossed by the
particles (grammage), see Ref.~\cite{2001ApJ...547..264J} and references of this
paper for details on the WSM.  In this model the transport equation
can be written at a given kinetic energy per nucleon $E_k$, as:
\begin{eqnarray}\label{impl}
  \lefteqn{\frac{\mathrm{d} N_i (x, E_k)}{\mathrm{d} x} =  \sum_T \frac{R_T (x)}{m_T} \times }\\\nonumber
  & & \Bigg\{ - \sigma_{iT} (E_k)  N_i (x, E_k) %\\\nonumber
 + \sum_{j>i} \sigma_{iT}^j N_j (x, E_k) \Bigg\},
\end{eqnarray}
where $N_i$ is the abundance of nuclear element $i$ at atmospheric depth $x$, $m_T$ being the target mass, 
$\sigma_{iR}$ the total reaction cross section and $\sigma_i^j$ the fragmentation cross section from a
nucleus $j$ to a (lighter, $i<j$) nucleus $i$. The TOA flux from ref~\cite{1998A&A...330..389W} were solar modulated and 
used for the initial conditions $N_i (0)$.
The sum over T corresponds to the various components of the atmosphere, with mass $m_T$ and fraction 
$R_T (x) = \frac{\rho_T(x)}{\rho_{\mathrm{tot}}(x)}$, which does not change significantly for the 
altitudes between ground and 200 km for the three main constituents \cite{1991JGR....96.1159H}. 
%
%\subsection{Model}\label{NUM}
%
The numerical approach of the problem has to be handled with care. The
inversion of equation~\ref{impl} where the TOA flux has to be computed
from the measured values leads to numerical difficulties and lengthy
calculations if the direct integration method is used. In this work,
the numerical calculations were instead, performed using the simpler,
easier to handle, matrix formulation of the problem, in which the
inversion is easy to achieve.

In this framework, the transport equation~\ref{impl} can be expressed as:
\begin{equation} \label{matr}
  \frac{\mathrm{d}\tilde{N}(x,E_k)}{\mathrm{d} x}= S(x,E_k) \tilde{N}(x,E_k) 
\end{equation}
where $\tilde{N}(x,E_k)$ is the vector containing all the elemental abundances
$N_i (x, E_k)$ of the considered CR flux for a traversed grammage of
$x$~g/cm$^2$ in the atmosphere. $S$ is the transformation matrix. It is a
triangular matrix, with the diagonal elements of $S$ corresponding to the
nuclear absorption and the other elements to the production of secondary nuclei.
The solution of \ref{matr} is given by:
\begin{equation} \label{sol}
  \tilde{N}(x,E_k)= R(x,E_k) \tilde{N}(0,E_k) 
\end{equation}
where $R(x,E_k) = \exp{(S(x,E_k))}$ is the transfer (transport) matrix.
%
%    \subsection{Nuclear cross sections}\label{CS}
%

To compute the $S$ and $R$ matrix, the parametrization from~\cite{TR99} was used
for the total reaction cross sections $\sigma_{iT}$, while the fit formula from
\cite{1990PhRvC..41..566W} was used for the fragmentation cross sections $\sigma_{iT}^j$. In
this latter case however, the formula applied mainly to H and He targets. Its
application was extended here to larger masses using a factorization
approximation~\cite{OL83}.
 
%   \subsection{Atmosphere}
%
The atmosphere model from \cite{1991JGR....96.1159H} was used in these calculations.

 \subsection{Results}
\begin{figure}[h!]
\begin{center}
\includegraphics [width=0.48\textwidth]{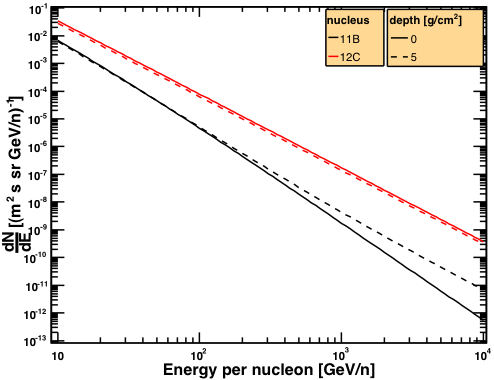}
\includegraphics [width=0.48\textwidth]{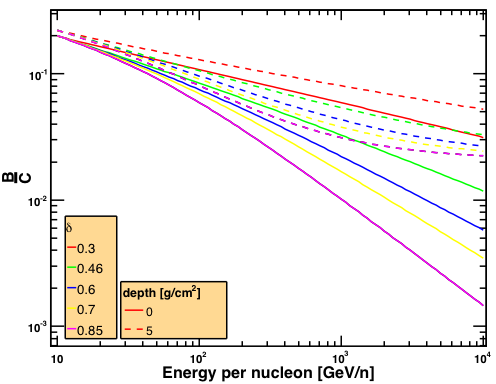}
\end{center}
\caption{Results of the transport calculations for C and B nuclei.
Top: Differential fluxes with $\delta = 0.3$ at TOA (solid lines) and
for a crossed matter thickness of 5 g/cm$^2$ (dashed lines) for
$^{12}$C (upper curves) and $^{11}$B (lower curves).  Bottom: B/C
ratio at TOA (solid lines) and after propagation (dashed lines) for
$\delta$: 0.3, 0.46, 0.6, 0.7, 0.85, from top to bottom respectively.
All curves as a function of the kinetic energy per
nucleon. \label{atmores}}
\end{figure}

Figure \ref{atmores} shows the non-propagated and propagated cosmic ray fluxes from a calculation 
%including $^{12}$C and $^{11}$B as primary CR nuclei with $^{11}$B being also produced by fragmentation 
%of $^{12}$C, 
for a spectral index of the diffusion coefficient $\delta = 0.3$, versus the energy per nucleon and 
%for the TOA flux of $^{11}$B (black line) and $^{12}$C (red line) and the propagated fluxes of $^{11}$B 
%(black dashed line) and $^{12}$C (red dashed line) at a grammage of 5 g/cm$^2$ of the Earth's atmosphere. 
%The energy dependence of the $^{12}$C propagated flux between 10 GeV/n and 10 Tev/n appears to be parallel 
%to the TOA flux for this nucleus, since it was assumed to be purely primary for the sake of simplicity. 
%To the opposite, the propagated flux of $^{11}$B becomes larger than the TOA flux at high 
%energy. This could be expected since the production through the fragmentation of the parent nucleus $^{12}$C 
%becomes dominant at these energies.
the correspondant B/C ratio for different values of the spectral index
of the diffusion coefficient $\delta$ = 0.3, 0.46, 0.6, 0.7, and
0.85. It can be seen that at the highest energy considered, the flux
values at TOA (solid lines) and at the balloon detection altitude
after crossing 5 g/cm$^2$ of atmosphere (dashed lines), differ by
approximately a factor of two for small values of $\delta$, and by
more than one order of magnitude for the largest values. This is easy
to understand since for larger $\delta$ the galactic contribution at
high energy becomes dominated by the atmospheric contribution. The
asymptotic limit of this behavior can be observed with the propagated
ratio tending to flatten out at high energy above 1~TeV/n for large
values of $\delta$ where the galactic secondaries production becomes negligible.

This illustrative example shows how critical are the corrections to be applied to the
measured flux values in balloon experiments, and thus how important is a careful
study of CR transport in the atmosphere for a reliable results in TOA flux
evaluations from atmospheric measurements.  In the high energy region, the
measured raw B/C value appears to be about one order of magnitude larger than
the TOA flux value to be extracted from it. 
%This will result in a very
%large uncertainty on the evaluated TOA value for these energies.
%
%les sections c/s, atmosphere, effects of stat, error study, sont remplacées chacune par un paragraphe avec refs. 

%\subsection{TOA flux reconstruction}

The flux at TOA can be reconstructed from the flux measured at a given
thickness $x$ by inverting the equation~\ref{sol}:
\begin{displaymath}
  \tilde{N}(0,E_k)= R(x,E_k)^{-1} \tilde{N}(x,E_k) 
\end{displaymath}
But in the inversion process, distortions may be generated by the random
(statistical) fluctuations of the experimentally measured fluxes
$\tilde{N} (x,E_k)$. Correcting for these distortions is equivalent to the
unfolding \cite{CO98} of a measured spectrum from the detector
response, the role of the latter being played here by the transfer
function (inverted matrix). This effect has been shown to be
negligible for the grammage (~5 g/cm$^2$) considered in this study~\cite{antjeDA}.

%\subsection{Effects of the systematic uncertainties}\label{ERRST}

The uncertainties induced on the flux calculations by the experimental
uncertainties on the nuclear cross-sections  have been
estimated by the following way for three values of the spectral index of the
diffusion coefficient $\delta$=0.3, 0.6 and 0.85, covering the range of realistic
values \cite{2001ApJ...555..585M,2005APh....24..146C}:

A sample of one hundred transfer matrix $R^{err}(x,E_k)$ was generated by adding
randomly a $\pm 5\%$ systematic error to the total reaction cross sections
(diagonal elements), and $\pm 10\%$ to the fragmentation cross sections (off
diagonal elements) in $R(x,E_k)$.  The measured fluxes were calculated using the
error added matrix
\begin{displaymath}
\tilde{N}^{\mathrm{err}}(x,E_k) = R^{\mathrm{err}}(x,E_k)\tilde{N}(0,E_k),
\end{displaymath}
and the TOA fluxes were reconstructed by the inversion procedure using
the nominal matrix $R(x)$:
\begin{displaymath}
\tilde{N}^{\mathrm{err}}(0,E_k) = R^{-1}(x,E_k)\tilde{N}^{\mathrm{err}}(x,E_k),
\end{displaymath}
Then the B/C ratios were calculated for each energy $E_k$, and the
minimal and maximal values of B/C were searched in the 100 values
calculated with the error-added matrices, and used as an estimate of
the upper and lower limits of the uncertainties induced by the
systematic errors on cross sections.
\begin{figure}[h!]
\begin{center}
\includegraphics[width=0.48\textwidth]{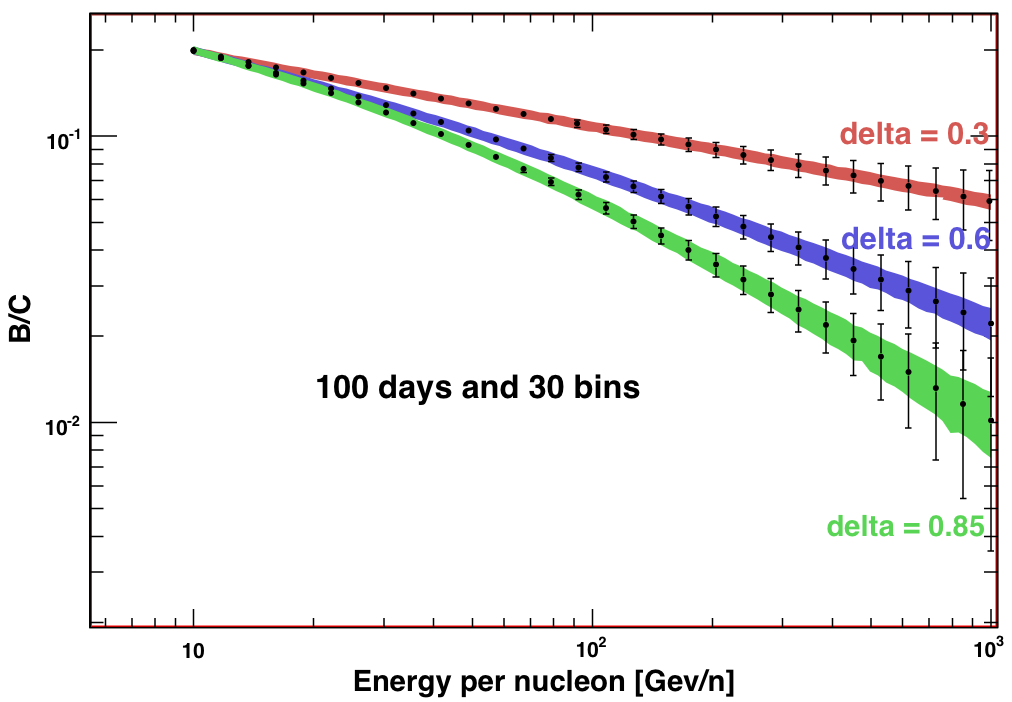}
\end{center}
\caption{The filled areas show the systematic error region obtained by the method
described in the text, for the three different values of $\delta$ =
0.3, 0.6, 0.85. The statistical error appear as the error bars on the
figure. They were calculated for 100 days of effective counting and
for 15 energy bins per decade of energy.} \label{error}
\end{figure}
Figure~\ref{error} shows the results from these calculations for three values of
the spectral index of the diffusion coefficient $\delta$=0.3, 0.6 and 0.85,
covering the range of realistic values \cite{2001ApJ...555..585M,2005APh....24..146C}. The values obtained
from the inversion of $R(x)$ are taken as central values. These results show
that as expected from fig~\ref{atmores}, for $\delta = 0.3$ the systematic error
is increasing with energy, but remains relatively small over the whole energy
range, in contrast with the $\delta = 0.85$ result, where it becomes 
large, extending over almost a factor 2 at high energy. As already mentioned, this can be easily
understood since for $\delta = 0.85$ more secondaries are produced in the
Earth's atmosphere compared to the secondary galactic flux compared with the
$\delta = 0.3$ productions (see figure \ref{atmores}). 

%Hence the error on the
%fragmentation cross sections induces a larger systematic error for $\delta =
%0.85$ than for $\delta = 0.3$. The large fragmentation cross sections
%measurements errors play a dominant role at high energies.

%For large $\delta$ values the cross section induced systematic errors
%do not allow making a clean discrimination between a $\delta = 0.8$
%and $\delta = 0.9$. For smaller values of $\delta$ however, it is
%possible to discriminate $\delta = 0.3$ from $\delta = 0.35$. The
%diffusive reacceleration model predicts a delta value between 0.5 and
%0.75. For these values an uncertainty of 0.07 is calculated.

Another question to be addressed is what counting statistics is needed for the
measurements to be dominated by systematics errors. A simple evaluation was made
by assuming a detector acceptance of 1~${m^2 \, sr}$, and taking 30 bins over
the energy range 10-1000~GeV/n, and  100 days of measurement time. The obtained
values are shown as error bars on the B/C ratios on Fig.~\ref{error} for the
three values of spectral indices $\delta$. It can be seen on the figure that the
statistical error is dominant for high energies ($E_k >$~200~GeV/n) for the
assumed experimental condition.
%\com{ça se voit comment ?}. 
A more quantitative study comparable to that reported in \cite{2005APh....24..146C}
remains to be performed.  In this reference, a $\chi^2$ minimization
procedure on the simulated B/C data was performed, to evaluate the
accuracy in the determination of $\delta$ due to statistical errors,
with all the propagation parameters left free. The estimated
statistical uncertainty on the determination of $\delta$ was 10-15\%
for experiments like CREAM.

\section{Summary and conclusion}\label{CONC}

The raw CR flux of nuclear elements measured in balloon experiments
must undergo significant corrections for the effects of propagation in
the atmosphere, to provide a reliable measurement of the TOA fluxes.
These corrections become larger with the increasing energy and with
increasing diffusion spectral index. Due to the uncertainties on the
absorption and fragmentation cross sections, they dramatically affect
the accuracy of the experimental values at high energy, but the measurements keep a
good part of their capacity to discriminate between the values of
$\delta$ predicted by the various models.

Since the same fragmentation process takes place in the detectors,
similar results can be expected, and the measured raw flux would have
to be corrected by a similar procedure as reported in this paper to
account for the flux propagation inside the detector material. The
correction will of course depend on the architecture of the apparatus
of the experiment and of the amount of matter upstream of the charge
measurement detector(s).

\section{Acknowledgments}

The authors are grateful to A. Castellina and F. Donato for making their
numerical results available to them.

%
%\newpage

%This is the reference to .bib file (Whitout .bib!)
\bibliography{icrc0888}
%This in the bibtex style, is ok.
\bibliographystyle{plain}
\end{document}